\documentclass{aa}
\usepackage{graphicx}
\usepackage{natbib}
\usepackage{amsmath}
\usepackage{amsfonts}
\bibpunct{(}{)}{;}{a}{}{,}

\begin{document}

\title{Comments on the paper \\ ``{\it {\bf The Mexican Hat Wavelet Family.
Application to \\ point source detection in CMB maps}}'' \\ by J. Gonz\'alez-Nuevo et. al (astro-ph/0604376)}

   \author{R. Vio\inst{1}
          \and
          P. Andreani\inst{2}
          }

   \offprints{R. Vio}

   \institute{Chip Computers Consulting s.r.l., Viale Don L.~Sturzo 82,
              S.Liberale di Marcon, 30020 Venice, Italy\\
              \email{robertovio@tin.it}
         \and
             INAF-Osservatorio Astronomico di Trieste,
             via Tiepolo 11, 34131 Trieste, Italy \\
		 Max-Planck f\"ur Extraterrestrische Physik, Postfach 1312, 85741 Garching, Germany \\
              \email{andreani@oats.inaf.it}
             }

\date{Received .............; accepted ................}

\abstract{The arguments presented by \citet{gon06} in favour of the {\it Mexican Hat Wavelet Family} (MHWF)
are critically discussed here. These authors allege the {\it optimal} properties of this new class of filters in 
the detection of point sources embedded in a noise background but their claim is not based upon
a solid mathematical foundation and proof.

}
\titlerunning{Optimal Detection of Sources}
\authorrunning{R. Vio, \& P. Andreani}
\maketitle

\section{Introduction}

The detection of (extragalactic) point sources embedded in a noise background is a critical issue
in the analysis of the experimental CMB maps. Actually, this kind of problem is not only
restricted to Cosmology. In fact,
this old dated and extensively studied question is of relevance in many scientific and engineering applications. 
Various approaches have been proposed. Among these, one of the most popular
is the {\it matched filter} (MF) technique. There are good reasons to choose it \citep[see ][]{kay98}:
\begin{itemize}
\item In the case of a background due to a Gaussian stochastic process (a rather common assumption),
MF is {\it optimal} in the Neyman-Pearson sense. This means that, for a given probability of {\it false alarm} 
(i.e., the probability of a spurious detection), it provides the best probability of detection; \\
\item If the Gaussian assumption is relaxed, MF still maximizes the {\it signal-to-noise ratio} (SNR)
at the output of {\it any} linear filter. In other words, among the linear filters, MF provides the greatest
enhancement of the magnitude of the source relative to the background;
\item As proved by its extensive and successful use in very
different contexts over many years, MF is a quite robust tool.
\end{itemize}
In spite of these remarkable characteristics, in the CMB literature a long series of papers were
published where the use of alternative techniques (e.g., {\it pseudo-filters}, {\it Mexican Hat Wavelet filters} and
{\it biparametric-scale adaptive filters}) is advocated and claimed to provide superior
results \citep[e.g., see ][]{san01, bar03, lop05}. In another series of papers
\citep{vio02, vio04, vio05a, vio05b} it was shown that such
claims were only the consequence of a incorrect interpretation of the results and/or of the
incorrect application of the validity conditions of some equations. Of course, this does not mean
that MF necessarily has to be chosen in every point source detection problems. However, given its
excellent properties, the use of alternative approaches must be motivated on the basis of well-grounded
theoretical arguments. Non-standard statistical tools are indicated only in situations
of real and sensible improvements of the results. New techniques that do not fulfill these requirements should
be introduced with care: they prevent the comparison with the results obtained in other works, may lead
people to use not well tested methodologies and end up in unreliable results.

Now, \citet{gon06} ({\it Go06}) seem do not agree on this commonsense rule since they propose a
new class of filters, say the {\it Mexican Hat Wavelet Family} (MHWF), that is
derived from the iterated application of the Laplacian operator to the {\it Mexican Hat Wavelet} (MHW).
However, they do not provide any formalized argument. Their claims are
supported only on numerical simulations. This is certainly not a safe way to proceed since,
contrary to a rigorous theoretical treatment, a set of numerical experiments is not sufficient to
characterize the statistical properties of a given methodology.

\section{The Mexican Hat Wavelet Filter vs. the Matched Filter}

{\it Go06} start from the consideration that, as written in their paper, {\it `` the Mexican 
Hat Wavelet is a very useful and powerful tool for point source detection due to the following reasons:
\begin{enumerate}
\item It has an analytical form that is very convenient when making calculations and that allows us to implement
fast algorithms. \\
\item It is well suited for the detection of Gaussian structure because it is obtained by applying the Laplacian
operator to the Gaussian function. \\
\item It amplifies the point sources with respect to the noise. Moreover, by changing the scale of the Mexican Hat
it is possible to control the amplification until an optimum value is achieved. \\
\item Besides, to obtain the optimal amplification it is not necessary to assume anything about the noise. In Vielva
et al. (2001), it was shown that the optimal scale can be easily obtained by means of a simple procedure for any
given image. Therefore the Mexican Wavelet is a very robust tool.''.
\end{enumerate}
}
Unfortunately, all of these points are irrelevant and/or false:
\begin{enumerate}
\item This property is overestimated. The computation of MF does not present
particular difficulties \citep[e.g., see ][]{vio02}. Moreover MHW and MF are linear filters, hence 
both of them can be implemented in very efficient algorithms; \\
\item It is not clear on the basis of which argument authors make this claim. Is there any theoretical
explanation for it? Moreover, if the conviction of authors is correct, why in the only analytical example
presented in their paper (i.e., Gaussian source plus white-noise background) the filter that provides the
best source amplification is not MHF but a Gaussian filter that, in this case, corresponds to MF? \\
\item Whatever filter that smoothes out the frequencies characteristic of the background
amplifies the point source with respect to the noise. But the best amplification is provided 
by MF not by MHW (see above).
This is true even in the case of the scale-adaptive version of MHF. In this respect, it is
necessary to stress a point that in \citet{gon06} is not adequately highlighted. In general, the performances
of MHW are by far inferior to those of MF
\citep[e.g., see ][]{vio02, vio04}. To make the scale $R$ a free parameter 
in the analytical form of MHF is only an attempt to improve a performance otherwise
unsatisfactory. The situation is similar to that of the fit of a set of data with an inadequate model.
If results are not good,  the simplest solution is to add one more free parameter to it.
Why one should choose MHW with a free parameter to be optimized 
when the only result that can be expected is a filter at the best suboptimal with respect to MF? \\
\item As written above, among all the linear filters, MF provides an optimal source amplification independently
from the nature of the noise. For this reason, it is not clear on which basis {\it Go06} claim that this property
belong to MHW.
\end{enumerate}
More in general, {\it Go06} plead that, given their ability to explore a signal at different
scales, the wavelets represent an effective tool in detection problems. However, it is not clear why this property 
should be useful in a context where all the sources have the same shape that, in addition, is known in advance. 
Why wavelets should provide superior results than the techniques, as MF, that make direct use of the shape of the 
sources? If a reason is available, this should be formally proved.

As last comment, we stress that MF can be easily adapted when dealing with sources with the
same shape but different amplitude. It it not difficult to prove that MF remains identical to
that corresponding to a single amplitude. Only the threshold determining the {\it probability of false alarm}
(hence also the probability of detection), is modified. If the distribution of the amplitudes is
known in advance, such threshold can be fixed through a Bayesian approach or, especially in the case of 
non-Gaussian background, through numerical simulations. 

\section{The Mexican Hat Wavelet Family}

By themselves, all the above points should be sufficient to question the real usefulness of a generalization of MHW.
However, let's assume for a moment that MHW has really optimal properties: which is the theoretical reason
to construct a family of filters through the iterated application of the Laplacian operator to MHW? Why
should such a procedure improve the performances of MHW? Quite surprisingly, {\it Go06} do not provide
any theoretical argument. The situation appears even more bizarre if one
considers again the only analytical example presented in their paper (Gaussian source plus white-noise background).
There, the source amplification worsens for increasing values of the order $n$. The {\it best} solution is
given by $n=0$. Since this case corresponds to the Gaussian filter, i.e. to MF, hopefully {\it Go06}
do not consider MF belonging to MHWF!

The ``optimal'' performances of MWHF is supported only
on the basis of the CMB numerical experiments carried out
by authors. We stress that this is insufficient
to drawn reliable conclusion about MHWF also because no whatever comparison is made with other filters
and specifically with MF. 

\section{Conclusions}

In this brief note the procedure suggested by \citep{gon06} for the detection of
point sources in CMB maps is criticized. The introduction of a new class of filters,
the {\it Mexican Hat Wavelet Family}, is given without providing any theoretical argument about
their real properties and usefulness. We stress that any new proposed statistical methodology
cannot be validated only on the basis of some numerical experiments, especially if no comparison with classic
and well tested techniques is made.

The risk of an uncontrolled proliferation of algorithms (and papers) whose reliability is at least dubious
must be avoided and we strongly hope that a more careful check on the scientific foundations will be
operated in the future.

\end{document}